# Anisotropic Positive Magnetoresistance of a Nonplanar 2D Electron Gas in a Parallel Magnetic Field

A. V. Goran[1], A. A. Bykov[1,*], A. K. Bakarov[1], and J. C. Portal[2]

[1]*Institute of Semiconductor Physics, Siberian Division, Russian Academy of Sciences, Novosibirsk, 630090 Russia*
*e-mail: bykov@thermo.isp.nsc.ru*
[2] *Grenoble High Magnetic Fields Laboratory, MPI-FKF and CNRS B.P.166, F-38042 Grenoble, France*

We study the transport properties of a 2D electron gas in narrow GaAs quantum wells with AlAs/GaAs superlattice barriers. It is shown that the anisotropic positive magnetoresistance observed in selectively doped semiconductor structures in a parallel magnetic field is caused by the spatial modulation of the 2D electron gas.

In an idealized zero-thickness 2D electron system, the orbital motion of charge carriers is affected only by the normal component of the external magnetic field, where the magnitude of this component depends on the angle between the magnetic field $\mathbf{B}_{ext}$ and the normal to the plane of 2D electron gas. The in-plane component of magnetic field in such a system will cause changes in the spin degree of freedom of charge carriers and, hence, in the density of states of 2D electron gas. The real 2D semiconductor systems always have a nonzero thickness, and this is the cause of the orbital effect in a parallel magnetic field [1]. Unlike the magnetoresistance associated with the spin effect [2], the one caused by the finite thickness of 2D electron gas is anisotropic. The origin of this anisotropy is that the variation of the effective mass of charge carriers in the direction perpendicular to the external magnetic field is greater than the variation in the direction parallel to the field.

This anisotropy mechanism manifests itself in the dependence of the magnetoresistance of 2D electron gas on the mutual orientation of the in-plane magnetic field and the measuring current. In particular, in the situation where the measuring current is perpendicular to the in-plane magnetic field, the magnetoresistance of 2D electron gas in AlGaAs/GaAs heterojunctions is greater than in the situation where the current is parallel to the field [3]. The anisotropy of positive magnetoresistance observed in [3] was found to be much smaller than that predicted by the theory [1]. In our opinion, this discrepancy is due to the fact that 2D electron gas in real selectively doped structures not only has a finite thickness but is also nonplanar [4–7]. As will be shown below, even a very small spatial modulation of 2D electron gas, which is inherent in any real structure, also leads to the anisotropy of the positive magnetoresistance of 2D electron gas in an in-plane magnetic field. However, the magnetoresistance in this mechanism is smaller when the magnetic field and the measuring current are mutually perpendicular and greater when they are parallel. A combined effect of the finite thickness and the spatial modulation of 2D electron gas should lead to a decrease in the degree of magnetoresistance anisotropy in the in-plane magnetic field, which may qualitatively explain the experimental results obtained in [3].

In the general case, the surface of 2D electron gas can be described by the function $z=z(x, y)$ characterizing the deviation of the surface from the ideal plane formed by the $x$ and $y$ axes. If we decompose the vector of external magnetic field into perpendicular and parallel components, $\mathbf{B}_{ext}=\mathbf{B}_\perp(x, y) + \mathbf{B}_\parallel(x, y)$, the quantities $\mathbf{B}_\perp$ and $\mathbf{B}_\parallel$ will be functions of $x$ and $y$. The perpendicular and parallel components are meant as the projections onto the normal vector and the tangential plane, respectively, at the point $(x, y)$ of the surface of 2D electron gas.

This decomposition is helpful because, in the case of a narrow quantum well, 2D electrons perceive only the normal component that is responsible for to the appearance of classical Larmor orbits in the plane of 2D electron gas. This normal component can be considered effective inhomogeneous magnetic field $B_{eff}(x, y)$ arising as a result of applying external magnetic field to the nonplanar 2D electron gas. In the particular case of the external magnetic field parallel to the sample, the effective field will be a sign alternating function with zero mean: $\langle B_{eff}(x, y)\rangle = 0$ [8]. The effective magnetic field $B_{eff}=B_{eff}(x, y)$ can be calculated if we know the surface of 2D electron gas $z=z(x, y)$ and the external magnetic field $\mathbf{B}_{ext}=(B_x, B_y, B_z)$. Then, $B_{eff}(x, y) = |B_{ext}|\cos(\Theta(x, y))$, where $|B_{ext}|$ is the magnitude of the vector of external magnetic field and $\Theta(x, y)$ is the angle between the normal to the surface $z=z(x, y)$ and the vector of external magnetic field $\mathbf{B}_{ext}$.

To characterize the surface of 2D electron gas, it is convenient to introduce the autocorrelation function $G(x, y) = \int (x - X, y - Y) z(x, y) dX dY$. If the surface is isotropic, the autocorrelation function will also be isotropic. However, the effective magnetic field $B_{eff}$ will be anisotropic, because the external magnetic field introduces a preferential direction into the system. Thus, the analysis of the influence of the in-plane magnetic field on the 2D electron transport in heterostructures with nonplanar heteroboundaries reduces to the problem of transport in an inhomogeneous magnetic field with zero mean and a certain anisotropy. Hence, in the general case, the magnetoresistance of a nonplanar 2D electron gas is due to the transport in an inhomogeneous magnetic field and should depend on the angle between the direction of the measuring current and the vector of the in-plane magnetic field.

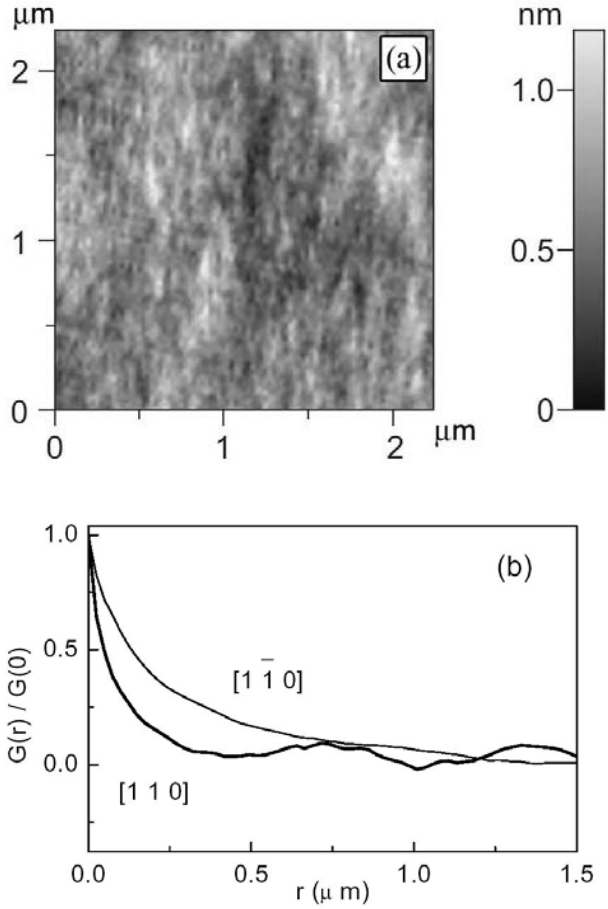

**Fig. 1.** (a) Two-dimensional AFM image of the surface of the MBE structure. (b) Autocorrelation functions of the surface relief in the [110] and [1$\bar{1}$0] directions.

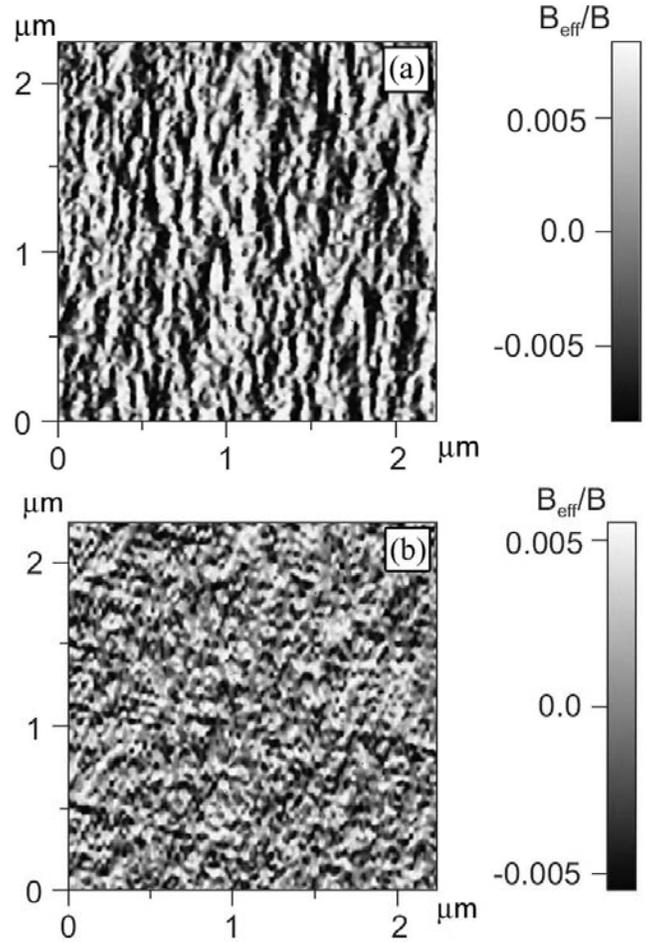

**Fig. 2.** Function $B_{eff}(x, y)$ calculated for a nonplanar 2D electron gas with a relief corresponding to the AFM image of the surface of MBE structure: $\mathbf{B}_{ext}$ is directed along (a) [110] and (b) [1$\bar{1}$0].

The structures studied in the experiment were selectively doped 10-nm-thick GaAs quantum wells with AlAs/GaAs superlattice barriers. They were prepared by molecular beam epitaxy (MBE) on (100) GaAs substrates whose deviation from the (100) plane did not exceed 0.02°. The surface morphology of the structures was examined by atomic force microscopy (AFM). Figure 1a shows the typical AFM image of the surface relief of the MBE structures under study. From the correlation analysis presented in Fig. 1b, one can see that the surface relief of a real MBE structure is not isotropic. The surface is wavy with a preferred [1$\bar{1}$0] orientation of the wave crests, which is typical of the selectively doped GaAs quantum wells with AlAs/GaAs superlattice barriers grown on GaAs (100) substrates [4].

Figure 2a shows a two-dimensional image of the effective magnetic field calculated for 2D electron gas on the assumption that its surface is identical to the AFM image of the MBE structure under study and that the external magnetic field is parallel to the [110] direction. One can clearly see that the effective magnetic field is anisotropic. The results of calculation of the effective magnetic field for the situation where the external magnetic field is oriented along the [1$\bar{1}$0] direction are shown in Fig. 2b. In this situation, the anisotropy of effective magnetic field is much smaller than in the previous case. Hence, for the 2D electron gas in the selectively doped MBE structures under study, the character of $B_{eff}(x, y)$ anisotropy depends on the direction of the vector $\mathbf{B}_{ext}$.

The magnetotransport experiments were carried out at temperatures from 4.2 to 1.6 K in magnetic fields up to 15 T on L-shaped Hall bars (Fig. 3a), which were fabricated by optical lithography and liquid etching. The bars had a width of 50 μm, and the distance between the potential terminals was 100 μm. The bar orientations were chosen so that the measuring current was parallel to the [110] and [1$\bar{1}$0] directions. The structures under study had one filled size-quantization subband. The equilibrium parameters of the 2D electron gas at $T = 4.2$ K were as follows: the concentration $n_s = 1.6 \times 10^{12}$ cm$^{-2}$ and the mobility $\mu = 300 \times 10^3$ cm$^2$/V s. Figure 3b represents the results of measurements of the relative magnetoresistance for two different orientations of the parallel external magnetic field: along the $x$ axis ($B_x = B_{ext}$ and $B_y = 0$) and along the $y$ axis ($B_x = 0$ and $B_y = B_{ext}$). Due to the anisotropy of the surface relief, the pattern of effective magnetic field for each of these $\mathbf{B}_{ext}$ directions is different. This results in the four combinations of the 2D electron-gas magnetoresistance along the [110] and [1$\bar{1}$0] directions. It should be noted that, in the temperature range from 4.2 to 1.6 K, the positive magnetoresistance observed in the structures under study did not vary, evidencing its classical [9, 10] rather than quantummechanical nature [11–13].

The anisotropy observed for the magnetoresistance of 2D electron gas can be qualitatively explained by electron scattering from the anisotropic inhomogeneous magnetic field [14] that depends on the angle between the vector $\mathbf{B}_{ext}$ and the direction of the measuring current. For the quantitative evaluation of this assumption, we carried out numerical simulation of the quasi-classical charge-carrier transport in the effective inhomogeneous magnetic field appearing in a nonplanar 2D electron gas in the parallel magnetic field. The model surfaces of the 2D electron gas were constructed using the results of AFM studies of real samples, for which the magnetic-field dependences are shown in Fig. 3b. The conductivity was calculated by the formula [15]

$$\sigma_{ij} = \frac{ne^2}{m} \int_0^\infty \langle v_i(0) v_j(t) \rangle e^{-t/\tau} dt,$$

where $v(t) = (\cos \varphi(t), \sin \varphi(t))$ is the direction of the electron velocity vector and $\tau$ is the charge-carrier transport relaxation time. The averaging was over $10^6$ trajectories. The factor $e^{-t/\tau}$ reflects the presence of impurities from which electrons are elastically scattered.

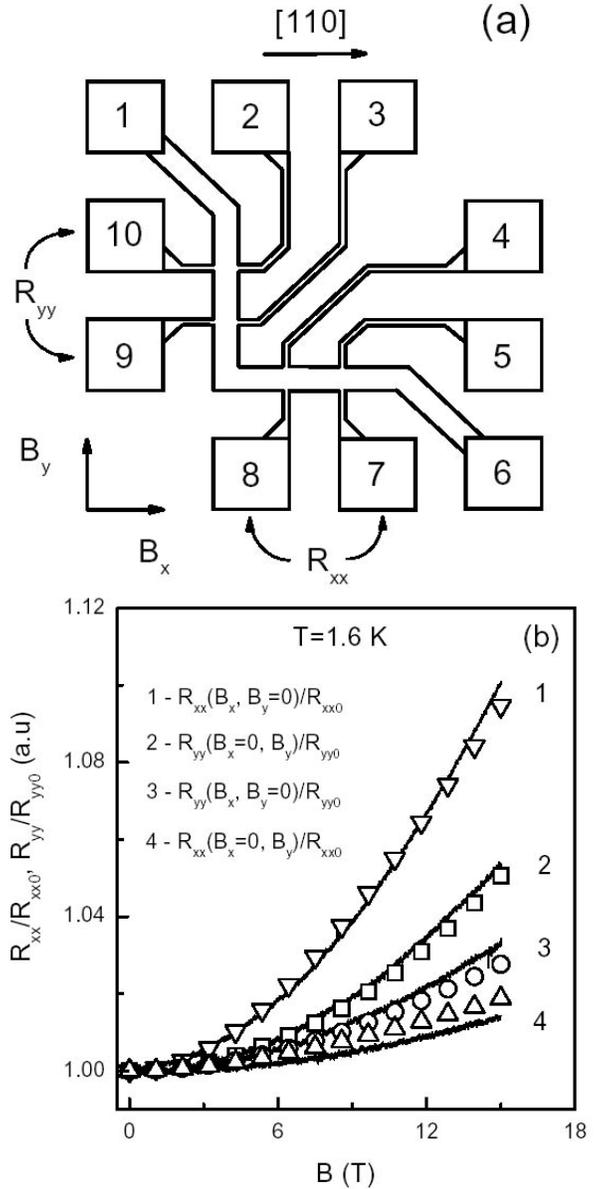

**Fig. 3.** (a) Schematic representation of an L-shaped Hall bar. (b) Dependences of the relative magnetoresistance of the 2D electron gas: (1) $R_{xx}(B_x, B_y = 0)/R_{xx0}$, (2) $R_{yy}(B_x = 0, B_y)/R_{yy0}$, (3) $R_{yy}(B_x, B_y = 0)/R_{yy0}$, and (4) $R_{xx}(B_x = 0, B_y)/R_{xx0}$. The solid lines represent the experimental curves, and the dots represent the calculations.

The trajectory of an electron $\mathbf{r}(t)$ was determined by the numerical integration of the equation of motion of a 2D electron in magnetic field:

$$m\dot{v}(t) = -\frac{e}{c}[v \times B_{eff}(x, y)],$$

where $\mathbf{B}_{eff}(x, y)$ is the effective magnetic field directed normally to the sample. By analogy with the experimental configurations, two directions were preset for the external field: along and across the [110] direction. In this way, four dependences were obtained for different combinations of $R_{xx}$, $R_{yy}$ and $B_x$, $B_y$. The model parameters were taken to be equal to the parameters of the real samples (mobility, concentration, and surface relief). The only fitting parameter was the amplitude of the spatial modulation of the 2D electron gas.

The results of modeling are shown in Fig. 3b. It should be noted that the calculated amplitude of spatial modulation of the 2D electron gas proved to be 2.5 times greater than the amplitude of surface roughness obtained from the AFM studies. We explain this difference by the fact that the 2D electron gas in the MBE structure under study is at a certain distance from the sample surface, and this distance is much greater than the roughness amplitude. Therefore, in the general case, the spatial modulation of the 2D electron gas may not coincide with the surface relief. One can see that the model and experimental dependences are in good agreement with this value of the fitting parameter.

Another possible explanation of the aforementioned discrepancy is that our calculations did not take into account the influence of the finite thickness of 2D electron gas on the value of positive magnetoresistance [1]. However, we believe that the high concentration of 2D electron gas and small width of the GaAs quantum well allow us to ignore the contribution from the orbital effect to the magnetoresistance. This approximation agrees with the absence of the temperature dependence of magnetoresistance in the interval from 4.2 to 1.6 K and allows us also to exclude other quantum-mechanical mechanisms from the consideration [11–13]. The quasi-classical nature of the magnetoresistance anisotropy observed by us is confirmed not only by the functional agreement between the model and experimental curves but also by the quantitative coincidence of the relative values of magnetoresistance obtained for different combinations of the directions of measuring current and parallel magnetic field. This result allows the following conclusion to be drawn: our model adequately describes the 2D electron transport in the selectively doped MBE structures under study, and the main contribution to the magnetoresistance comes from the scattering by the effective inhomogeneous magnetic field arising in such structures in a parallel external magnetic field.

Thus, we have shown that the anisotropic positive magnetoresistance of a high-concentration 2D electron gas in a parallel magnetic field is governed by the scattering by the effective inhomogeneous magnetic field, i.e., by the spatial modulation of the 2D electron gas in the selectively doped MBE structures under study.

This work was supported by the Russian Foundation for Basic Research, project no. 04-02-16789.